\documentclass[10pt,preprint]{aastex}

\usepackage{float}

\shorttitle{$\Sigma$-$D$ Relation for Shell-Type SNRs}
\shortauthors{Hu et al.}

\begin{document}


\title{Theoretical $\Sigma$-$D$ Relations for Shell-Type Galactic Supernova Remnants}


\author{Ya-Peng Hu\altaffilmark{*,1,2}, Hong-An Zeng\altaffilmark{\natural,1}, Jun Fang\altaffilmark{\dag,3}, Jun-Peng Hou\altaffilmark{\ddag,1}, and Jian-Wen Xu\altaffilmark{\sharp,2}}





\altaffiltext{*,$\natural$,$\dag$,$\ddag$,$\sharp$}{$*$Email:huyp@nuaa.edu.cn;$\natural$hazeng@nuaa.edu.cn;$\dag$fangjun@ynu.edu.cn; $\ddag$jack.hjp@gmail.com;$\sharp$xjw@itp.ac.cn}
\altaffiltext{1}{College of Science, Nanjing University of Aeronautics and Astronautics, Nanjing 210016, China}
\altaffiltext{2}{Key Laboratory of Frontiers in Theoretical Physics, Institute of Theoretical Physics, Chinese Academy of Sciences, Beijing 100080, China}
\altaffiltext{3}{Department of Astronomy, Yunnan University, Kunming 650091, China}


\begin{abstract}
Relations between radio surface brightness ($\Sigma$) and diameter ($D$) of supernova remnants (SNRs) are important in astronomy. In this paper, following the work Duric \& Seaquist (1986) at adiabatic phase, we carefully investigate shell-type supernova remnants at radiative phase, and obtain theoretical $\Sigma$-$D$ relation at radiative phase of shell-type supernova remnants at 1 GHz. By using these theoretical $\Sigma$-$D$ relations at adiabatic phase and radiative phase, we also roughly determine phases of some supernova remnant from observation data.

\end{abstract}






\section{Introduction}

Relations between radio surface brightness ($\Sigma$) and diameter ($D$) of supernova remnants (SNRs) are important in astronomy, and are usually used to determine distance of a SNR (Poveda \& Woltjer 1968; Clark \& Caswell 1976;
Lozinskaya 1981; Huang \& Thaddeus 1985; Duric \& Seaquist 1986;
Guseinov et al. 2003). There have been many works via statistical or analytical approaches to investigate $\Sigma$-$D$ relations (e.g., Poveda \& Woltjer 1968; Clark \& Caswell
1976; Mills et al. 1984; Huang \& Thaddeus 1985;
Arbutina et al. 2004, Xu et al. 2005, Pavlovic et al. 2014, etc.). Among statistical results of $\Sigma$-$D$ relations, one straight line is often
obtained by authors (e.g., Poveda \&
Woltjer 1968; Huang \& Thaddeus 1985; Arbutina et al. 2004; Pavlovic et al. 2013; Pavlovi et al. 2014), while a broken fit line or a transition point is also usually seen. For example, Clark
\& Caswell (1976), Allakhverdiyev et al. (1983), and Allakhverdiyev
et al. (1985) have gotten a broken fit line in their statistical works. At 408~MHz, Clark
\& Caswell (1976) had a broken line with slopes of $\beta =
-2.7/-10$ ($\Sigma \propto D^{\beta}$) at $D\leq 32$~pc/$D\geq
32$~pc, while Allakhverdiyev et al. (1983)
got $30$~pc at 408~MHz and $32$~pc at 1~GHz for 15 shell-type remnants. For analytical $\Sigma$-$D$ relations, Duric \& Seaquist (1986) once derived
\begin{equation}\label{SigmaD11}
  \Sigma(D)=4\times10^{-14}D^{-5},     D\ll1pc
\end{equation}
\begin{equation}\label{SigmaD12}
  \Sigma(D)=4\times10^{-15}D^{-3.5},   D\gg1pc
\end{equation}

On the other hand, galactic supernova remnants are usually classified into three types: Shell-type, Plerion-type and Composite-type. In our paper, for simplicity, we just focus on investigating shell-type galactic supernova remnants. Usually, shell-type galactic supernova remnants have four evolution stages: the free expansion phase, adiabatic or Sedov phase, radiative or snowplough phase and the dissipation phase. Nearly all of detected shell-type SNRs are at adiabatic phase or radiative phase, since almost none is observed in the 1st and 4th phases due to fact that shell-type SNRs at these two phases are usually practically undetectable. Therefore, $\Sigma$-$D$ relations of shell-type supernova remnants at adiabatic phase and radiative phase are interesting issues to investigate. Indeed,  Duric \& Seaquist (1986) have analytical derived $\Sigma$-$D$ relation of shell-type supernova remnants at adiabatic phase, i.e. the above equation (\ref{SigmaD12}). In our paper, we will mainly focus on analytical investigations on $\Sigma$-$D$ relations of shell-type supernova remnants. We obtain theoretical $\Sigma$-$D$ relation at radiative phase of shell-type supernova remnants at 1 GHz, which is simply followed the work Duric \& Seaquist (1986). We also have collected 57 shell-type galactic supernova remnants data where some data have been updated according to Green (2004, 2009 \& 2014) and other new references. By using these theoretical $\Sigma$-$D$ relations at adiabatic phase and radiative phase, we have roughly determined the phases of some supernova remnant among these 57 shell-type galactic supernova remnants data.

This article is organized as follows. In section 2, after a brief review of the work Duric \& Seaquist (1986) at adiabatic phase, we simply follow their work and further analytically investigate $\Sigma$-$D$ relation at radiative phase of shell-type galactic supernova remnants at 1 GHz. In section 3, we have collected 57 shell-type galactic supernova remnants data. By using these two theoretical $\Sigma$-$D$ relations at adiabatic phase and radiative phase, we roughly determine the phases of some supernova remnant among these data. Finally, a brief conclusion and discussion are given in section~\ref{discuss}.

\section{Theoretical $\Sigma$-$D$ Relations at adiabatic phase and radiative phase}\label{theory}

In this section, we mainly focus on theoretical $\Sigma$-$D$ relations of shell-type supernova remnants. After making a brief review on the analytical work Duric \& Seaquist (1986) at adiabatic phase, we theoretically derive $\Sigma$-$D$ relation at radiative phase of shell-type supernova remnants.

\subsection{A brief review : work Duric \& Seaquist (1986) }\label{sedov}


Taking the linear diameter ($D$) of remnant in pc, time ($t$) in s, SNR initial
explosion energy ($E_{0}$) in the unit of $10^{51}$ ergs with $E_{0}=\varepsilon_{51}\times 10^{51}$, and ISM electron
density ($n_0$) in cm$^{-3}$, from the standard Sedov solution, one
has the following equation
(Bignami \& Caraveo 1988, Zaninetti 2000, V\"olk et al.
2002, Ptuskin \& Zirakashvili 2003)
\begin{equation}\label{Dt2}
D(t) = A_0 t ^{2/5},
\end{equation}
where the coefficient is
\begin{equation}\label{A02}
A_0 =  5.4\times 10^{-4} \left( \begin{array}{c} \frac{\varepsilon_{51}}{n_0}
\end{array} \right) ^{1/5}.
\end{equation}
The shock wave velocity should be
\begin{equation}\label{vt2}
\upsilon (t) = \frac{1}{2}\frac{d}{dt}D(t) = \frac{1}{5} A_0 t ^{-3/5}.
\end{equation}
At the adiabatic phase, the thickness of remnant is
proportional to $D$, and the shell volume
which contains all the radio-emitting particles is
\begin{equation}\label{VD2}
V(D) = C_0 D ^3,
\end{equation}
here $C_0 = \frac{\pi}{6} ( 1-(\frac{D_i}{D_0})^3)\simeq 0.37$
is the volume coefficient (Milne 1970). Notice that condition $D_i/D_o\sim 2/3$ has
been assumed and $D_i$ and $D_o$ are the
inner and outer diameter of the remnant shell respectively. Combining (\ref{Dt2})
 and (\ref{VD2}), one obtains the volume of the shell with respect to $t$
\begin{equation}\label{Vt2}
V(t) = C_0 A_0^3 t^{6/5}.
\end{equation}
As the shock waves of remnant travel, the ambient magnetic field $B$
at the adiabatic phase will decrease with $D$ according to (Duric \& Seaquist 1986)
\begin{equation}\label{BD2}
B(D) = B_0 \left( \begin{array}{c} \frac{D_0}{D} \end{array} \right)
^{2}.
\end{equation}
Substituting (\ref{Dt2}) to it, we have
\begin{equation}\label{Bt2}
B(t) = B_0 D_0^2 A_0^{-2} t^{-4/5}.
\end{equation}
Ginzburg \& Synovatskii (1965) and Bell (1978) show that the radio
emissivity $\epsilon(B, \upsilon)$ of a shocked gas which are
affected by a magnetic field to produce the synchrotron emission
is expressed as (Arbutina B. et al. 2012)
\begin{eqnarray}\label{eBv2}
&\epsilon(\nu) = 2.94\times 10^{-34} (1.435 \times 10^5)^{0.75 -
\alpha} \xi(2\alpha + 1)\nonumber\\
&\times \left(\frac{n_0}{cm^{-3}} \right) \left(
\frac{\alpha}{0.75} \right)
\left(\frac{\upsilon}{10^4~km~s^{-1}} \right)^{4\alpha}  \left(
\frac{B}{10^{-4} G} \right) ^{\alpha + 1}\nonumber\\
&\times  \left( 1 + \left( \frac{\upsilon}{7000~Km~s^{-1}} \right) ^{-2} \right) ^{\alpha}
\left( \frac{\nu}{GHz} \right) ^{-\alpha},  \label{Radio emissivity}
\end{eqnarray}
where $\psi_e=4$ and $\phi_e=10^{-3}$ have been set as same as in Bell (1978), and unit of $\epsilon(\nu)$ is $W Hz^{-1} m^{-3}$, $\xi(\mu) = 11.7 a(\mu)$, and $a(\mu)$ is the function
tabulated by Ginzburg \& Synovatskii (1965). The velocities
 of shock waves in the second and third phase of SNRs are
typically far less than $7000~Km~s^{-1}$.
Thus, (\ref{Radio emissivity}) can be further simplified
\begin{eqnarray}
\epsilon (\nu )&=&2.94\times 10^{-34}\times (1.435\times
10^{5})^{0.75-\alpha }\xi (2\alpha +1)\nonumber\\
&&\times(\frac{\alpha }{0.75})(0.7)^{4\alpha }(\frac{n_0}{cm^{-3}})(%
\frac{v}{7000km/s})^{2\alpha }(\frac{B}{10^{-4}G})^{\alpha
+1}(\frac{\nu }{GHz})^{-\alpha }\nonumber\\
&=&2.94\times 10^{-34}\times (1.435\times
10^{5})^{0.75-\alpha }\xi (2\alpha +1)\nonumber\\
&&\times(\frac{\alpha }{0.75})(0.7)^{4\alpha }(\frac{n_0}{cm^{-3}})(%
\frac{v}{2.3\times 10^{-10}pc/s})^{2\alpha }(\frac{B}{10^{-4}G})^{\alpha
+1}(\frac{\nu }{GHz})^{-\alpha }\nonumber\\
&&~(WHz^{-1}m^{-3}).\label{Radio emissivity2}
\end{eqnarray}
Note that, in the second line of this equation, shock velocity $v$ in unit of $pc/s$ has been considered for the later convenience of discussion. Taking account of (\ref{Dt2}), (\ref{Bt2}) and the
average value of the remnants spectral index $\alpha = 0.5$, we can
get
\begin{eqnarray}\label{eD2}
\epsilon(D) = 2.25\times 10^{-34} \left(\begin{array}{c}
\frac{D_0}{D} \end{array} \right)^3 \left(
\begin{array}{c} \frac{B_0}{10^{-4} G} \end{array} \right) ^{3/2}\nonumber\\
\times \left( \begin{array}{c} \frac{1}{5} A_0^{5/2} D^{-3/2}/(2.3\times 10^{-10}pc/s)
 \end{array} \right).
\end{eqnarray}
which is at 1 GHz and $a(2)=0.103$ has been used in Ginzburg \& Synovatskii (1965). If the shell volume is considered to be encompassed by the radiating
electrons, the surface brightness of remnant can be written as (Duric \& Seaquist 1986)
\begin{eqnarray}\label{Sigmat2}
\Sigma(t) = \frac{\epsilon(t)V(t)}{\pi^2 D^2(t)}.
\end{eqnarray}
Inserting (\ref{Dt2})~(\ref{A02})~(\ref{Vt2}) and (\ref{eD2}) into it, we obtain
\begin{eqnarray}\label{SigmaD21}
\Sigma(D) = 2.25\times 10^{-34} \frac{C_0 D_0^3}{\pi^2 D^2}
\left(\begin{array}{c} \frac{B_0}{10^{-4} G} \end{array}
\right) ^{3/2}\nonumber\\
\times \left( \begin{array}{c} \frac{1}{5} A_0^{5/2} D^{-3/2}/(2.3\times 10^{-10}pc/s) \end{array} \right).
\end{eqnarray}
Finally, one can get
\begin{equation}\label{SigmaD22}
\Sigma(D) = m_a D_{pc}^{-3.5}~ (W m^{-2} Hz^{-1} sr^{-1}),
\end{equation}
where
\begin{eqnarray}\label{ma2}
m_a &=& 2.25\times 10^{-34} \frac{C_0 D_0^3}{\pi^2}
\left( \frac{B_0}{10^{-4} G} \right) ^{3/2}
\times \left( \begin{array}{c} \frac{1}{5} A_0^{5/2}/(2.3\times 10^{-10}pc/s) \end{array} \right) \times 3.08 \times 10^{16}\nonumber\\
&=&3.88\times 10^{-17},
\end{eqnarray}
and $\Sigma(D)$ in (\ref{SigmaD21}) in unit of $(W m^{-3} Hz^{-1}~pc~sr^{-1})$ and $1 pc= 3.08 \times 10^{16} m$ have been considered, and some typical values of physical parameters of SNRs are taken:
ISM density $n_0 =0.1$~cm$^{-3}$, SNR initial explosion energy $E_0
=10^{51}$~erg, the diameter and ISM magnetic field of remnant at the
beginning of Sedov phase $D_0 =2$~pc and $B_0 =10^{-4}$~G, etc.
Therefore, the analytically derived line of $\Sigma$-$D$ relation
at the second phase of shell-type SNR is
\begin{eqnarray}\label{SigmaD2l}
\Sigma(D) = 3.88\times 10^{-17} D_{pc}^{-3.5}
~(W m^{-2} Hz^{-1} sr^{-1}).\label{Relation at second phase}
\end{eqnarray}
Note that, different typical values have been chosen, and hence coefficient in~(\ref{Relation at second phase}) is a little different from that in work Duric \& Seaquist 1986, but the power-law reminds the same exponent.



\subsection{Analytical $\Sigma$-$D$ relation at the radiative phase}\label{Rad}
It should be pointed out that the above work Duric \& Seaquist (1986) just analytically investigate the adiabatic phase of shell-type SNRs. In fact, we can also simply follow their work to analytically investigate $\Sigma$-$D$ relation at radiative phase of shell-type SNRs. After setting the same choices of units as those in section (\ref{sedov}), the
equation for shell-type SNRs at the radiative stage is (Mckee et al. (1977))
\begin{equation}\label{Dt3}
D(t) = A_1 t ^{2/7},
\end{equation}
where $A_1$ is a constant
\begin{equation}\label{A03}
A_1 = 0.03 \left( \begin{array}{c} \frac{\varepsilon_{51}}{n_0}
\end{array} \right) ^{1/7}.
\end{equation}
From which, we obtain the velocity of shock wave at the radiative phase
\begin{equation}\label{vt3}
\upsilon (t) = \frac{1}{7} A_1 t ^{-5/7}.
\end{equation}
Same as the adiabatic phase, the volume of shell can be
\begin{equation}\label{VD3}
V(D) = C_1 D ^3.
\end{equation}
If we roughly take $D_i/D_o\sim 3/4$, then the coefficient will be
$C_1 = \frac{\pi}{6} (1-(\frac{D_i}{D_0})^3) \simeq 0.3$. Changing the variant
$D$ to $t$, one can rewrite the volume of shell as
\begin{equation}\label{Vt3}
V(t) = C_1 A_1^3 t^{6/7}.
\end{equation}
Note that the ambient magnetic field $B$ of a remnant decreases with the diameter $D$ at the adiabatic phase following equation (\ref{BD2}),
while at the dissipation-phase it is $B(D) = B_1(D_1/D)^0$.
Therefore, we can assume that the ambient magnetic field
$B$ at the radiative phase can be expressed as
\begin{equation}\label{BD3}
B(D) = (\frac{D_1}{D})^\beta B_1,
\end{equation}
where the parameter $\beta$ ranges from 0 to 2. After
substituting (\ref{Dt3}) to it, one gets
\begin{equation}\label{Bt3}
B(t) = \left( \frac{D_1}{A_1} \right)^\beta B_1 t^{-2\beta/7}.
\end{equation}
Therefore, following the same steps as the above section and still taking $n_0 = 0.1$~cm$^{-3}$, we can obtain $\Sigma-D$
relation at radiative phase at 1 GHz
\begin{eqnarray}\label{SigmaD31}
\Sigma(D) &=& 2.25\times 10^{-34} \frac{C_1 D^3}{\pi^2 D^2}
\left(\begin{array}{c} \frac{B_1 D_1^\beta D^{-\beta}}{10^{-4}
G} \end{array} \right) ^{3/2}\nonumber\\
&\times& \left( \begin{array}{c} \frac{1}{7} A_1^{7/2} D^{-5/2}/(2.3\times 10^{-10}pc/s) \end{array} \right).
\end{eqnarray}
and this form is simply rewritten as
\begin{equation}\label{SigmaD32}
\Sigma(D) = m_r D^{-\frac{3}{2}(1+\beta)}~ (W m^{-2} Hz^{-1} sr^{-1}),
\end{equation}
where
\begin{eqnarray}\label{mr3}
m_r &=& 2.25\times 10^{-34} \frac{C_1 D_1^{\frac{3}{2}\beta}}{\pi^2}
\left(\begin{array}{c} \frac{B_1}{10^{-4}
G} \end{array} \right) ^{3/2}\nonumber\\
&\times& \left( \begin{array}{c} \frac{1}{7} A_1^{7/2}/(2.3\times 10^{-10}pc/s) \end{array} \right)\times 3.08 \times 10^{16}.
\end{eqnarray}

\section{Roughly determine phases of some supernova remnants by using theoretical $\Sigma$-$D$ relations}

For these theoretical $\Sigma$-D relations at adiabatic and radiative phases, it will be highly interesting to identify some reasonable supernova remnants in adiabatic or radiative stages from observation data. Therefore, at first, some typical values of supernova remnants at radiative phase are also set as $B_1 = 10^{-6}$G, $D_1 = 20$ pc, $\beta = 1$. Then, theoretical $\Sigma$-D relation at radiative phase is
\begin{eqnarray}
\Sigma(D) = 1.86 \times 10^{-16} D_{pc}^{-3}~(W m^{-2} Hz^{-1} sr^{-1}). \label{StdInit}
\end{eqnarray}
Second, 57 shell-type supernova remnants data in Galaxy at 1 GHz have been collected and listed in table~\ref{57dtd}. From these data, the 1GHz surface brightness $\Sigma_{1GHz}$ is obtained by  (Clark \& Caswell 1976)
\begin{eqnarray}
\Sigma_{1GHz}=1.505 \frac{S_{1GHz}}{\theta^2}\times 10^{-19}~(W m^{-2}Hz^{-1} sr^{-1}),
\end{eqnarray}
where $S_{1GHz}$ is the 1GHz flux density in jansky ($1 Jy\equiv10^{-26}Wm^{-2}Hz^{-1}$), and $\theta$ is the angular diameter in minutes of arc.

Finally, comparing $\Sigma_{1GHz}$ with theoretical $\Sigma$-D relations both at adiabatic phase and radiative phase, we roughly determine phases of supernova remnants. From these comparisons, we find out that some of these supernova remnants indeed can be identified in adiabatic phase or radiative phase, which has been listed in table~\ref{Phase-information}.

\newpage

\begin{table}[!h]
\centering
\begin{tabular}{cccccccc}
\multicolumn{8}{c}{Determine adiabatic phase or radiation phase}\\
\hline
Source &   Dia. &  $\Sigma$ &  $\Sigma_0$ &  Deviation &   $\Sigma_1$ &   Deviation &   state\\
\hline
G8.7$-$0.1      &51   &5.95E$-$21   &4.11E-23   &99.31\%    &1.40E-21   &76.42\%      &radiative\\
G32.8$-$0.1     &35   &5.73E$-$21   &1.53E-22   &97.32\%    &4.34E-21   &24.27\%      &radiative\\
G33.6$+$0.1     &23   &3.01E$-$20   &6.67E-22   &97.79\%    &1.53E-20   &49.21\%      &radiative\\
G41.1$-$0.3     &8    &2.97E$-$20   &2.69E-20   &9.64\%     &3.63E-19   &-1122.00\%   &adiabatic\\
G43.3$-$0.2     &10   &3.97E$-$20   &1.23E-20   &69.03\%    &1.86E-19   &-368.33\%    &adiabatic\\
G49.2$-$0.7     &52   &2.68E$-$20   &3.84E-23   &99.86\%    &1.32E-21   &95.06\%      &radiative\\
G78.2$+$2.1     &26   &1.34E$-$20   &4.34E-22   &96.76\%    &1.06E-20   &20.89\%      &radiative\\
G109.1$-$1.0    &24   &4.22E$-$21   &5.74E-22   &86.40\%    &1.35E-20   &-218.59\%    &adiabatic\\
G111.7$-$2.1    &5    &1.32E$-$19   &1.39E-19   &-5.68\%    &1.49E-18   &-1029.95\%   &adiabatic\\
G127.1$+$0.5    &69   &8.91E$-$22   &1.43E-23   &98.40\%    &5.66E-22   &36.51\%      &radiative\\
G132.7$+$1.3    &51   &1.06E$-$21   &4.11E-23   &96.12\%    &1.40E-21   &-32.51\%     &radiative\\
G205.5$+$0.5    &102  &4.35E$-$22   &3.63E-24   &99.17\%    &1.75E-22   &59.74\%      &radiative\\
G284.3$-$1.8    &20   &2.87E$-$21   &1.09E-21   &62.17\%    &2.33E-20   &-708.94\%    &adiabatic\\
G327.4$+$0.4    &29   &1.02E$-$20   &2.96E-22   &97.11\%    &7.63E-21   &25.51\%      &radiative\\
G327.6$+$14.6   &19   &3.18E$-$21   &1.30E-21   &59.05\%    &2.71E-20   &-753.50\%    &adiabatic\\
G330.0$+$15.0   &63   &1.63E$-$21   &1.96E-23   &98.79\%    &7.44E-22   &54.25\%      &radiative\\
G332.4$-$0.4    &9    &4.21E$-$20   &1.78E-20   &57.79\%    &2.55E-19   &-505.47\%    &adiabatic\\
G337.8$-$0.1    &27   &9.29E$-$22   &3.80E-22   &59.06\%    &9.45E-21   &-917.19\%    &adiabatic\\
G346.6$-$0.2    &19   &1.88E$-$20   &1.30E-21   &93.08\%    &2.71E-20   &-44.15\%     &radiative\\
G349.7$+$0.2    &9    &1.20E$-$19   &1.78E-20   &85.23\%    &2.55E-19   &-111.91\%    &adiabatic\\
\hline
\end{tabular}
\caption{$\Sigma$ is directly obtained from experimental data in Table\ref{57dtd}. For each supernova remnant with observational diameter D, $\Sigma_0$ is obtained from theoretical $\Sigma$ -$D$ relation at adiabatic phase, while $\Sigma_1$ is from theoretical $\Sigma$ -$D$ relation at radiative phase. After making comparisons with $\Sigma$, and obtaining the deviations (i.e., deviation of $\Sigma_0$ is just simply calculated from $(\Sigma-\Sigma_0)/\Sigma$), we roughly determine supernova remnant at adiabatic phase or radiation phase. } \label{Phase-information}
\end{table}

\section{Conclusion and Discussion}\label{discuss}


In this paper, we have analytically investigated $\Sigma$-$D$ relations of shell-type supernova remnants both at adiabatic phase and radiative phase. For convenience to compare with observation data, we have chosen some typical values of shell-type supernova remnants and also collected 57 shell-type supernova remnants data. By using these theoretical $\Sigma$-$D$ relations and observation data, we have roughly identified some shell-type supernova remnants in adiabatic phase or radiative phase.

Some discussions related to our results are in the following. First, shock compression ratio may differ among supernova remnants, and hence spectral index $\alpha$ will be different from $\alpha=0.5$. For simplicity and convenience, we also investigate the case with $\alpha=0.75$, and theoretical $\Sigma$-$D$ relation at adiabatic phase
\begin{eqnarray}
\Sigma(D) = 1.06 \times 10^{-17} D_{pc}^{-19/4}~(W m^{-2} Hz^{-1} sr^{-1}), \label{StdInitnew1}
\end{eqnarray}
while at radiative phase
\begin{eqnarray}
\Sigma(D) = 5.52 \times 10^{-16} D_{pc}^{-9/2}~(W m^{-2} Hz^{-1} sr^{-1}). \label{StdInitnew2}
\end{eqnarray}
From these equations, it will be easily found that effects from spectral index $\alpha$ on $\Sigma$-$D$ relations are in fact huge, i.e. different power exponents. Second, not only spectral index $\alpha$, our results are also dependent on other parameters such as volume coefficient $C_0$, mean electron density $n_0$, SNR initial explosion energy $E_0$, magnetic
field at the beginning of the evolving second-stage and third-stage
$B_0$ and $B_1$, and parameters $D_0$ and $D_1$. If those parameters are changed, our results may be also different. Therefore, further effects of parameters on theoretical $\Sigma$-$D$ relations will be an interesting open issue, while maybe they are also the main reason that why just some of supernova remnants are roughly identified at adiabatic phase or radiative phase among 57 supernova remnants, and hence our results just shed some insights onto the possibility to identify the phase of supernova remnant through comparisons theoretical $\Sigma$-$D$ relations with observation data. Third, the true physical process of supernova remnant at radiative phase is complicated, and we have assumed that the synchrotron radiation equation~(\ref{Radio emissivity2}) is still valid in radiative stage. But yet, whether this assumption is correct or not is still an issue (e.g., discussions in Asvarov 2006). Comparison with observational data seem to support this assumption. Fourth, in principle, there is a simple and direct method to identify the phase of supernova remnant, i.e. comparisons $D(t)$ relations with observation data. However, this method may be not better than our method, and the reason may be that there is a larger uncertainty to decide the age of a supernova remnant than diameter. Finally, our results also predict that there will be a transition point between these two theoretical $\Sigma$-$D$ relations. However, since $D(t)$ relations used in our paper are just statistical and not precise, the details of this transition point are still lacking.


\section{Acknowledgments}
Y.P Hu thanks Profs. Hongsheng Zhang and Nan Liang for useful discussions and Dr. Ai-Yuan Yang for providing some updated data, and also thanks anonymous referee for constructive and helpful comments. This work is supported by the National Natural Science Foundation of China (NSFC) under grants No.11575083, 11565017, 11105004, the Fundamental Research Funds for the Central Universities under grant No. NS2015073, and Shanghai Key Laboratory of Particle Physics and Cosmology under grant No. 11DZ2260700. Jun Fang is supported by NSFC No. 11563009 and Yunnan Applied Basic Research Projects No. C0120150289.

\clearpage

\clearpage

\begin{deluxetable}{lrrrrcrrrrr}
\tablewidth{0pt}
\tablecaption{Some physical parameters of 57 shell-type Galactic SNRs.\label{57dtd}}
\tablehead{
\colhead{Source}           & \colhead{Age/year}      &
\colhead{Dist./pc}          & \colhead{Dia./pc}  &
\colhead{size/arcmin}          & \colhead{S$_{1GHz}$/Jy}    &
\colhead{ref}}
\startdata
 G4.5$+$6.8 &  380     &   2900      &  3     &  3        &  19     &  H90, G04a\\
   G7.7$-$3.7 &  $-$     &   4500      &  29    &  22       &  11     &  M86      \\
   G8.7$-$0.1 &  15800   &   3900      &  51    &  45       &  80     &  G96      \\
  G18.8$+$0.3 &  16000   &   12000     &  57    &  17x11    &  33     &  G04a, TL07 \\
  G27.4$+$0.0 &  2700    &   6800      &  8     &  4        &  6      &  G04a, C82\\
  G31.9$+$0.0 &  4500    &   7200      &  13    &  7x5      &  25     &  CS01, G14 \\
  G32.8$-$0.1 &  $-$     &   7100      &  35    &  17       &  11     &  K98b     \\
  G33.6$+$0.1 &  9000    &   7800      &  23    &  10       &  20     &  S03, SV95, G04a, G14\\
  G39.2$-$0.3 &  1000    &   11000     &  22    &  8x6      &  18     &  G14, C82  \\
  G41.1$-$0.3 &  1400    &   8000      &  8     &  4.5x2.5  &  25     &  C99, C82, B82, G14\\
  G43.3$-$0.2 &  3000    &   10000     &  10    &  4x3      &  38     &  L01, ZT14      \\
  G49.2$-$0.7 &  30000   &   6000      &  52    &  30       &  160    &  KKS95, G04a\\
  G53.6$-$2.2 &  15000   &   2800      &  24    &  33x28    &  8      &  S95, G04a\\
  G55.0$+$0.3 &  1100000 &   14000     &  71    &  20x15   &  0.5    &  MWT98    \\
  G65.3$+$5.7 &  14000   &   1000      &  78    &  310x240  &  42     &  G14, R81\\
  G73.9$+$0.9 &  10000   &   1300      &  8     &  27      &  9      &  LLC98, G14, L89\\
  G74.0$-$8.5 &  14000   &   400       &  23    &  230x160  &  210    &  LGS99, SI01, G04a\\
  G78.2$+$2.1 &  50000   &   1500      &  26    &  60       &  320    &  LLC98, KH91, G14\\
  G84.2$-$0.8 &  11000   &   4500      &  23    &  20x16    &  11     &  MS80, M77, G04a\\
  G89.0$+$4.7 &  19000   &   800       &  24    &  120x90   &  220    &  LA96     \\
  G93.3$+$6.9 &  5000    &   2200      &  15    &  27x20    &  9      &  L99, G04a\\
  G93.7$-$0.2 &  $-$     &   1500      &  35    &  80       &  65     &  UKB02    \\
 G109.1$-$1.0 &  17000   &   4000      &  24    &  28       &  22     &  FH95, G04a, G14, HHv81, TL12\\
 G111.7$-$2.1 &  320     &   3400      &  5     &  5        &  2720   &  TFv01    \\
 G114.3$+$0.3 &  41000   &   700       &  15    &  90x55    &  5.5      &  MBP02, G04a, G14\\
 G116.5$+$1.1 &  280000  &   1600      &  32    &  80x60    &  10     &  G04a, G14, RB81\\
 G116.9$+$0.2 &  44000   &   1600      &  16    &  34       &  8      &  KH91, G04a, G14\\
 G119.5$+$10.2&  24500   &   1400      &  37    &  90      &  36     &  M00      \\
 G120.1$+$1.4 &  410     &   2300      &  5     &  8        &  56     &  H90, G04a\\
 G127.1$+$0.5 &  85000   &   5250      &  69    &  45       &  12     &  G14, FRS84 \\
 G132.7$+$1.3 &  21000   &   2200      &  51    &  80       &  45     &  G04a, GTG80\\
 G156.2$+$5.7 &  26000   &   2000      &  64    &  110      &  5      &  RFA92    \\
 G160.9$+$2.6 &  7700    &   1000      &  38    &  140x120  &  110    &  LA95     \\
 G166.0$+$4.3 &  81000   &   4500      &  57    &  55x35    &  7      &  KH91, G04a, L89\\
 G166.2$+$2.5 &  150000  &   8000      &  186   &  90x70    &  11     &  G14, RLV86    \\
 G182.4$+$4.3 &  3800    &   3000      &  44    &  50       &  0.4    &  KFR98, G14\\
 G205.5$+$0.5 &  50000   &   1600      &  102   &  220      &  140    &  CB99, G14     \\
 G206.9$+$2.3 &  60000   &   7000      &  102   &  60x40    &  6      &  G14, L86      \\
 G260.4$-$3.4 &  3400    &   2200      &  35    &  60x50    &  130    &  B94, RG81\\
 G266.2$-$1.2 &  680     &   1500      &  52    &  120      &  50     &  K02, AIS99\\
 G272.2$-$3.2 &  6000    &   1800      &  8     &  15      &  0.4    &  D97      \\
 G284.3$-$1.8 &  10000   &   2900      &  20    &  24      &  11     &  G14, RM86     \\
G296.5$+$10.0&  20000   &   2000      &  44    &  90x65    &  48     &  G14, MLT88    \\
 G296.8$-$0.3 &  1600000 &   9600      &  47    &  20x14    &  9      &  GJ95, G04a\\
 G299.2$-$2.9 &  5000    &   500       &  2     &  18x11    &  0.5    &  SVH96    \\
 G309.2$-$0.6 &  2500    &   4000      &  16    &  15x12    &  7      &  RHS01    \\
 G315.4$-$2.3 &  2000    &   2300      &  28    &  42       &  49     &  DSM01, G04a\\
 G321.9$-$0.3 &  200000  &   9000      &  70    &  31x23       &  13     &  G14, SFS89, S89\\
 G327.4$+$0.4 &  $-$     &   4800      &  29    &  21       &  30     &  SKR96, G04a, G14, WS88\\
 G327.6$+$14.6&  980     &   2200      &  19    &  30       &  19     &  G04a, SBD84\\
 G330.0$+$15.0&  $-$     &   1200      &  63    &  180     &  350    &  K96      \\
 G332.4$-$0.4 &  2000    &   3100      &  9     &  10       &  28     &  CDB97, G04a, MA86\\
 G337.2$-$0.7 &  3250    &   15000     &  26    &  6        &  1.5      &  RHS01, G14    \\
 G337.8$-$0.1 &  $-$     &   12300     &  27    &  9x6      &  18     &  K98b     \\
 G346.6$-$0.2 &  $-$     &   8200      &  19    &  8        &  8      &  K98b, D93\\
 G349.7$+$0.2 &  14000   &   11500     &  9     &  2.5x2    &  20     &  RM01, G04a, TL14\\
 G352.7$-$0.1 &  2200    &   8500      &  17    &  8x6      &  4      &  K98a     \\
\enddata
\tablenotetext{a}{Many of the radio SNRs have more than one
published value for distance and age. For these, we either chose the
most recent estimates or used an average of the available estimates,
or the most commonly adopted value.}
\tablenotetext{b}{Diameters were calculated using from distances together with the
angular sizes in Green (2004, 2009 \& 2014) catalogue. In addition, some data have been updated according to the new results in Green (2014) and other new references.}
\tablenotetext{c}{Some data regarding G349.7+0.2 (TL14), G43.3-0.2 (ZT14), G18.8+0.3 (TL07) and G109.1-1.0 (TL12) have been
updated.}

\end{deluxetable}





\begin{thebibliography}{}
\bibitem[Allakhverdiyev et al. (1983)]{alla83} Allakhverdiyev A.O., Amnuel P.R., Guseinov O.H., Kasumov F.K. 1983, Ap\&SS, 97, 287
\bibitem[Allakhverdiyev et al. (1985)]{alla85} Allakhverdiyev A.O., Guseinov O.H., Kasumov F.K., Yusifov I.M. 1985, Ap\&SS, 121, 21
\bibitem[Arbutina et al. (2004)]{arbu04} Arbutina B., Urosevic D., Stankovic M., Tesic Lj. 2004, MNRAS, 350, 346
\bibitem[Arbutina B et al. (2012)]{atbu14} Arbutina B., Urosevic D., Andjelic M.M., Pavlovic M.Z., Vukotic B., 2012, ApJ, 746, 79
\bibitem[Aschenbach B et al.(1999)]{asch99} Aschenbach B., Iyudin A.F., Sch\"onfelder V. 1999, A\&A, 350, 997 (AIS99)
\bibitem[Bell(1978)]{bell78} Bell, A. R. 1978, MNRAS, 182, 443
\bibitem[Berthiaume et al. (1994)]{bert94} Berthiaume G.D., Burrows D.N., Garmire G.P., Nousek J.A. 1994, ApJ, 425, 132 (B94)
\bibitem[Bignami et al.(1988)]{bign88} Bignami, G. F., P. A. Caraveo, and J. A. Paul. 1988, A\&A, 202, L1-L4
\bibitem[Binette et al.(1982)]{bine82} Binette L., Dopita M.A., Dodorico S., Benvenuti P. 1982, A\&A, 115, 315 (B82)
\bibitem[Carter et al.(1997)]{cart97} Carter L.M., Dickel J.R., Bomans D.J. 1997, PASJ, 109, 990 (CDB97)
\bibitem[Clark \& Caswell(1976)]{clar76} Clark D.H., Caswell J.L. 1976, MNRAS, 174, 267
\bibitem[Case \& Bhattacharya(1999)]{case99} Case G.L., Bhattacharya D. 1999, ApJ, 521, 246 (CB99)
\bibitem[Caswell et al.(1982)]{casw82} Caswell J.L., Haynes R.F., Milne D.K., Wellington K.J. 1982, MNRAS,
    200, 1143 (C82)
\bibitem[Chen \&Slane(2001)]{chen01} Chen Y., Slane P.O. 2001, ApJ, 563, 202 (CS01)
\bibitem[Chen et al.(1999)]{chen99} Chen Y., Sun M., Wang Z.R., Yin Q.F. 1999, ApJ, 520, 737 (C99)
\bibitem[Dickel et al.(2001)]{dick01} Dickel J.R., Strom R.G., Milne D.K. 2001, ApJ, 546, 447 (DSM01)
\bibitem[Dubner et al.(1993)]{dubn93} Dubner G.M., Moffett D.A., Goss W.M., Winkler P.F. 1993, AJ, 105, 2251 (D93)
\bibitem[Duncan et al.(1997)]{dunc97} Duncan A.R., Stewart R.T., Campbell-Wilson D., Haynes R.F., Aschenbach B., Jones K.L. 1997, MNRAS, 289, 97 (D97)
\bibitem[Duric \& Seaquist(1986)]{duri86} Duric N., Seaquist E.R. 1986, ApJ, 301, 308
\bibitem[Fesen \& Horford(1995)]{fese95} Fesen R.A., Horford A.P. 1995, AJ, 110, 747 (FH95)
\bibitem[F\"urst et al.(1984)]{furs84} F\"urst E., Reich W., Steube R. 1984, A\&A, 133, 11 (FRS84)
\bibitem[Gaensler \& Johnston(1995)]{gaen95} Gaensler B.M., Johnston S. 1995, MNRAS, 277, 1243 (GJ95)
\bibitem[Galas et al.(1980)]{gala80} Galas C.M.F., Tuohy L.R., Garmire G.P. 1980, ApJ, 236, L13 (GTG80)
\bibitem[Ginzburg \& Syrovatskii(1965)]{ginz65} Ginzburg, V. L., Syrovatskii, S. I. 1965, Annual Review of Astronomy and Astrophysics 3 : 297
\bibitem[Gorham et al.(1996)]{gorh96} Gorham P.M., Ray P.S., Anderson S.B., Kulkarni S.R., Prince T.A. 1996, ApJ, 458, 257 (G96)
\bibitem[Green(2004)]{gree04} Green D.A. 2004, arXiv:astro-ph/0411083vl, 3 (G04a)
\bibitem[Green(2009)]{gree09} Green D.A. 2009, arXiv:0905.3699
\bibitem[(2014)]{gree14} Green D.A. 2014, arXiv:1409.0637v1 (G14)
\bibitem[(2003)]{guse03} Guseinov O.H., Ankay A., Sezer A., Tagieva S.O. 2003, A\&AT, 22, 273
\bibitem[(1990)]{hats90} Hatsukade I., Tsunemi H., Yamashita K., Koyama K., Asaoka Y., Asaoka I. 1990, PASJ, 42, 279 (H90)
\bibitem[(1985)]{huan85} Huang Y.L., Thaddeus P. 1985, ApJ, 295, L13
\bibitem[(1981)]{hugh81} Hughes V.A., Harten R.H., van den Bergh S. 1981, ApJ, 246, L127 (HHv81)
\bibitem[(2002)]{karg02} Kargaltsev O., Pavlov G.G., Sanwal D., Garmire G.P. 2002, ApJ, 580, 1060 (K02)
\bibitem[(1998)]{kinu98} Kinugasa K., Torii K., Tsunemi H., Yamauchi S., Koyama K., Dotani T. 1998, PASJ, 50, 249 (K98a)
\bibitem[(1996)]{knod96} Kn\"oedlseder J., Oberlack U., Diehl R., Chen W., Gehrels N. 1996, A\&AS, 120, 339 (K96)
\bibitem[(1991)]{koo91} Koo B.C., Heiles C. 1991, ApJ, 382, 204 (KH91)
\bibitem[(1995)]{koo95} Koo B.C., Kim K.T., Seward F.D. 1995, ApJ, 447, 211 (KKS95)
\bibitem[(1998)]{kora98} Koralesky B., Frail D.A., Goss W.M., Claussen M.J., Green A.J. 1998, ApJ, 116, 1323 (K98b)
\bibitem[(1998)]{koth98} Kothes R., F\"urst E., Reich W. 1998, A\&A, 331, 661 (KFR98)
\bibitem[(2001)]{lace01} Lacey C.K., Joseph T., Lazio W. Kassim N.E., Duric N., Briggs D.S.,
    Dyer K.K., 2001, ApJ, 559, 954 (L01)
\bibitem[(1999)]{land99} Landecker T.L., Routledge D., Reynolds S.P., Smegal R.J., Borkowski K.J., Seward F.D. 1999, ApJ, 527, 866 (L99)
\bibitem[(1986)]{leah86} Leahy D.A. 1986, A\&A, 156, 191 (L86)
\bibitem[(1989)]{leah89} Leahy D.A. 1989, A\&A, 216, 193 (L89)
\bibitem[(1995)]{leah95} Leahy D.A., Aschenbach B. 1995, A\&A, 293, 853 (LA95)
\bibitem[(1996)]{leah96} Leahy D.A., Aschenbach B. 1996, A\&A, 315, 260 (LA96)
\bibitem[(1999)]{leve99} Levenson N.A., Graham J.R., Snowden S.L. 1999, ApJ, 526, 874 (LGS99)
\bibitem[(1998)]{lori98} Lorimer D.R., Lyne A.G., Camilo F. 1998, A\&A, 331, 1002 (LLC98)
\bibitem[(1981)]{lozi81} Lozinskaya T.A. 1981, Soviet Astron. Lett., 7, 17
\bibitem[(1988)]{mats88} Matsui Y., Long K.S., Tuohy I.R. 1988, ApJ, 329, 838 (MLT88)
\bibitem[(1977)]{matt77} Matthews H.E., Baars J.W.M., Wendker H.J., Goss W.M. 1977, A\&A, 55, 1 (M77)
\bibitem[(1980)]{matt80} Matthews H.E., Shaver P.A. 1980, A\&A, 87, 255 (MS80)
\bibitem[(1998)]{matt98} Matthews B.C., Wallace B.J., Taylor A.R. 1998, ApJ, 493, 312 (MWT98)
\bibitem[(2002)]{mavr02} Mavromatakis F., Boumis P., Paleologou E.V. 2002, A\&A, 383, 1011 (MBP02)
\bibitem[(2000)]{mavr00} Mavromatakis F., Papamastorakis J., Paleologou E.V., Ventura J. 2000, A\&A, 353, 371 (M00)
\bibitem[(1977)]{mcke77} McKee, Christopher F., and Jeremiah P. Ostriker. 1977, AJ, 218, 148
\bibitem[(1986)]{meab86} Meaburn J., Allan P.M. 1986, MNRAS, 222, 593 (MA86)
\bibitem[(1984)]{mill84} Mills B.Y., Turtle A.J., Little A.G., Durdin J.M. 1984, Australian J. Phys., 37, 321
\bibitem[(1970)]{miln70} Milne D.K. 1970, Australian J. Phys., 23, 425
\bibitem[(1986)]{miln86} Milne D.K., Roger R.S., Kesteven M.J., Haynes R.F., Wellington K.J., Stewart R.T. 1986, MNRAS, 223, 487 (M86)
\bibitem[(2013)]{pavl13} Pavlovic M.Z., Urosevic D., Vukotic B., Arbutina B., and Goker U.D., 2013, Astrophys.J.Suppl., 204, 4
\bibitem[(2014)]{pavl14} Pavlovic M.Z., Dobardzic A., Vukotic B., and Urosevic D., 2014, Serb.Astron.J. 189, 25, arXiv:1411.2234
\bibitem[(1968)]{pove68} Poveda A., Woltjer L. 1968, AJ, 73, 65
\bibitem[(2003)]{ptus03} Ptuskin V.S., Zirakashvili V.N. 2003, A\&A, 403, 1
\bibitem[(2001)]{rako01} Rakowski C.E., Hughes J.P., Slane P. 2001, ApJ, 548, 258 (RHS01)
\bibitem[(1981)]{reic81} Reich W., Braunsfurth E. 1981, A\&A, 99, 17 (RB81)
\bibitem[(1992)]{reic92} Reich W., F\"urst F., Arnal E.M. 1992, A\&A, 256, 214 (RFA92)
\bibitem[(2001)]{reyn01} Reynoso E.M., Mangum J.G. 2001, ApJ, 121, 347 (RM01)
\bibitem[(1981)]{rosa81} Rosado M. 1981, ApJ, 250, 222 (R81)
\bibitem[(1981)]{rosa81} Rosado M. Gonz\'alez,J., 1981, Rev. Mexicana. Astron. Astrof., 5, 93 (RG81)
\bibitem[(1986)]{rout86} Routledge D., Landecker T.L., Vaneldik J.F. 1986, MNRAS, 221, 809 (RLV86)
\bibitem[(1986)]{ruiz86} Ruiz M.T., May J. 1986, ApJ, 309, 667 (RM86)
\bibitem[(1995)]{sake95} Saken J.M., Long K.S., Blair W.P., Winkler P.F. 1995, ApJ, 443, 231 (S95)
\bibitem[(1989)]{salt89} Salter C.J., Reynolds S.P., Hogg D.E., Payne J.M., Rhodes P.J. 1989,
    ApJ, 338, 171 (S89)
\bibitem[(1996)]{sewa96} Seward F.D., Kearns K.E., Rhode K.L. 1996, ApJ, 471, 887 (SKR96)
\bibitem[(2003)]{sewa03} Seward F.D., Slane P.O., Smith R.K., Sun M. 2003, ApJ, 584, 414 (S03)
\bibitem[(1995)]{sewa95} Seward F.D., Velusamy T. 1995, ApJ, 439, 715 (SV95)
\bibitem[(1989)]{shul89} Shull J.M., Fesen R.A., Saken J.M. 1989, ApJ, 346, 860 (SFS89)
\bibitem[(1996)]{slan96} Slane P., Vancura O., Hughes J.P. 1996, ApJ, 465, 840 (SVH96)
\bibitem[(1984)]{srin84} Srinivasan G., Bhattacharya D., Dwarakanath K.S. 1984, J, Astrophys. Astr.,
    5, 403 (SBD84)
\bibitem[(2001)]{stil01} Stil J.M., Irwin J.A. 2001, ApJ, 563, 816 (SI01)
\bibitem[(2001)]{thor01} Thorstensen J.R., Fesen R.A., van den Bergh S. 2001, AJ, 122, 297 (TFv01)
\bibitem[(2007)]{tian07} Tian W.W., Leahy D.A., Wang Q.D. 2007, A\&A, 474, 541 (TL07)
\bibitem[(2012)]{tian12} Tian W.W., Leahy D.A. 2012, MNRAS, 421, 2593 (TL12)
\bibitem[(2014)]{tian14} Tian W.W., Leahy D.A. 2014, ApJL, 783, L2 (TL14)
\bibitem[(2002)]{uyan02} Uyaniker B., Kothes R., Brunt C.M. 2002, ApJ, 565, 1022 (UKB02)
\bibitem[(2002)]{volk02} V\"olk H.J., Berezhko E.G., Ksenofontov L.T., Rowell G.P. 2002, A\&A, 396, 649
\bibitem[(1988)]{weil88} Weiler K.W., Sramek R.A. 1988, ARA\&A, 26, 295 (WS88)
\bibitem[(2005)]{xu05}   Xu J.W., Zhang X. Z., Han J. L. 2005, CJAA, 5, 165
\bibitem[(2000)]{zani00} Zaninetti L. 2000, A\&A, 356, 1023
\bibitem[(2014)]{zhu14} Zhu H., Tian W.W., Zuo P. 2014, ApJ, 793, 95 (ZT14)
\bibitem[(2006)]{Asvarov06} Asvarov A.I. 2006, A\&A, 459, 519


\end{thebibliography}
\end{document}